# Heart Rate Variability during Periods of Low Blood Pressure as a Predictor of Short-Term Outcome in Preterms

Oksana Semenova*, Giorgia Carra*, Gordon Lightbody, Geraldine Boylan,
Eugene Dempsey, Andriy Temko, *Senior Member, IEEE*

*Abstract*—Efficient management of low blood pressure (BP) in preterm neonates remains challenging with a considerable variability in clinical practice. The ability to assess preterm wellbeing during episodes of low BP will help to decide when and whether hypotension treatment should be initiated. This work aims to investigate the relationship between heart rate variability (HRV), BP and the short-term neurological outcome in preterm infants less than 32 weeks gestational age (GA). The predictive power of common HRV features with respect to the outcome is assessed and shown to improve when HRV is observed during episodes of low mean arterial pressure (MAP) - with a single best feature leading to an AUC of 0.87. Combining multiple features with a boosted decision tree classifier achieves an AUC of 0.97. The work presents a promising step towards the use of multimodal data in building an objective decision support tool for clinical prediction of short-term outcome in preterms who suffer episodes of low BP.

## I. Introduction

Preterm birth is a leading cause of neonatal mortality. Every year more than one in ten babies are born preterm and this number is rising [1]. Preterm babies are at a higher risk of complications, which may lead to short-term and long-term adverse health outcomes. Low blood pressure (BP) or hypotension is known to be one of main problems in preterms particularly during the first 72 hours after delivery. It may cause inadequate blood flow to the brain, heart and other vital organs. The definition of hypotension is still uncertain and management of preterm neonates with low BP remains challenging with significant variability in practice [2]. When the level of mean arterial pressure (MAP) falls below the gestational age (GA) in weeks, intervention in the form of blood volume expansion is commonly initiated. This threshold, however, lacks any published supporting evidence. At the same time hypotension treatment has been associated with adverse outcomes including brain injury [3]. For babies with no clinical or biochemical signs of shock, no intervention may well be appropriate.

The ability to assess preterm wellbeing during episodes of low BP may improve efficiency of hypotension management. ECG is routinely recorded in preterms from which the time variation between successive heartbeats (heart rate variability, HRV) can be estimated. HRV provides a non-invasive assessment of both the parasympathetic and sympathetic control of heart rate.

In term neonates, a significant association between HRV features, severity of hypoxic ischemic brain injury and long-term neurodevelopmental outcome at two years of age was reported for 61 full-term neonates in [4]. Promising results have been obtained for newborn outcome prediction using a combination of multimodal features including HRV [5]. In preterms, the high frequency component of HRV was shown to be a good biomarker of necrotizing enterocolitis, an acute neonatal inflammatory disease which may lead to death [6]. HRV features together with quantification of general physical movements have also been found useful for 2-year outcome prediction in [7]. In [8], time-domain HRV measurements have showed a significant difference between septic and non-septic newborns. Decreased HRV was observed in children born with low birth weight, including preterm babies [9]. Measurements of HRV have been assessed as a predictor for successful removal of mechanical ventilation in [10]. A correlation between low frequency oscillations of HRV and BP for preterm neonates has been reported in [11].

While several works have tried to study an association between HRV and neonatal health outcome in the term and preterm population, there is still a lack of understanding of this relationship for preterms in the context of low BP episodes.

This study aims to investigate the relationship between HRV, BP and the short-term neurological outcome in preterm infants who were less than 32 weeks GA. In particular, the usefulness of HRV for the prediction of outcome is assessed and the predictive power of HRV features during the episodes of low BP is studied.

## II. Materials and Methods

### A. Dataset

The dataset is comprised of 35 preterm infants with mean GA of 28 weeks (range: 23 – 31 weeks) recorded at the Neonatal Intensive Care Unit of Cork University Maternity Hospital, Ireland.

The aim of this study is to investigate the effect of low BP on the preterm wellbeing and the threshold was chosen to be quite loose, GA+4 (as compared to GA). Twenty three

This research was supported by a Science Foundation Ireland Research Centers Award (SFI 12/RC/2272).

O. Semenova, G. Lightbody, A. Temko are with the Department of Electrical and Electronic Engineering, Irish Center for Fetal and Neonatal Translational Research (INFANT), University College Cork, Ireland. (e-mail: o.semenova@umail.ucc.ie, {atemko, g.lightbody}@ucc.ie)

G. Boylan and E. Dempsey are with the Department of Pediatrics and Child Health, INFANT center, UCC, Ireland:{g.boylan, g.dempsey}@ucc.ie.

G. Carra is with the Laboratory of Intensive Care Medicine, KU Leuven, Belgium, giorgia.carra@kuleuven.be. The work was performed during her stay at UCC.

*These authors contributed equally to this work.

preterm neonates were selected out of 35 (mean GA is 27 weeks) based on the criteria that during the recording there was at least one episode of at least 5 min duration where MAP fell below a threshold (MAP < GA +4). The data set included continuous ECG and simultaneous registration of BP. The total duration of the 23 recordings is 824 hours. A ViasysNicOne video EEG machine (CareFusion Co., San Diego, USA) was used to record ECG sampled at either 256 Hz or 1024 Hz. Continuous invasive arterial BP monitoring was simultaneously performed via an umbilical arterial catheter using the Philips Intellevue MP70 machine, which provides BP data sampled at 1 Hz. Positioning of the tip of the umbilical catheter in the descending aorta was confirmed by chest radiograph.

Short-term neurological outcome of the preterm was represented by clinical course score (CCS) which was assigned independently to every infant by two consultant neonatologists [12]. CCS is a binary score, which is based on the discharge summary and medical notes summarising the presence or absence of five major neonatal complications: grade III/IV intraventricular haemorrhage or cystic periventricular leukomalacia, necrotizing enterocolitis, bronchopulmonary dysplasia, infection and retinopathy. The study had full ethical approval from the Clinical Research Ethics Committee of the Cork Teaching Hospitals. In the cohort of 23 preterms, 12 had healthy outcome.

### B. Signal processing and feature extraction

The diastolic (DP) and systolic (SP) pressures were recorded every second and used to calculate the mean arterial pressure: $MAP = DP + 1/3\,(SP - DP)$.

The ECG signal was segmented into non-overlapping 5-minute epochs. The R-peaks were identified using the Pan-Tompkins method [13]. Abnormal values of time intervals between R-peaks (RR intervals) caused by artefacts were corrected by moving average where possible. Periods of clear movement artefact were automatically discarded. The RR intervals were used to estimate the heart rate HRV signal. In order to explore the clinical value of HRV, the RR intervals were normalized with respect to the average RR interval, obtaining the normal-to-normal interval (NN). In this study fifteen time and frequency domain features were extracted from the NN and RR intervals. These features have been previously used for various purposes to quantify the HRV in term and preterm infants [5], [14].

The time domain features include the statistics of the RR interval distribution and include mean (meanRR), standard deviation (SDNN), skewness, kurtosis, triangular interpolation (TINN).

Other features include the temporal information, such as root mean square of successive NN intervals (RMSSD) and the ratio between SDNN and RMSSD. The Poincare diagram is used to graphically describe beat-to-beat variability of the RR intervals and is constructed by plotting every RR interval against the preceding one. The main dimensions SD1 and SD2 of the Poincare diagram were used to quantify beat-to-beat variations and a continuous shift in the RR interval.

Nonlinear HRV features include approximate entropy (ApEn) to quantify regularity and complexity of stationary signal and the Allan Factor (AF) which quantifies the variability of successive counts.

Frequency domain features were obtained from uniformly resampled NN intervals, at 256 Hz. HRV was then quantified by power various bands: very low frequency (0.008 – 0.04 Hz) band (VLF); low frequency (0.04 – 0.2 Hz) band (LF), high frequency (0.2 - 1 Hz) band (HF) as well as the ratio between the LF and HF power (LF/HF) [5].

### C. Statistical analysis: feature predictive power

The area under the receiver operating characteristic curve (AUC) is used to quantify the discriminative (predictive) power of each HRV feature with respect to health status of preterm, which is represented by CCS. The AUC is directly connected to the Mann-Whitney U-statistics which is a robust non-parametric alternative to the student's t-test. An area of 0.5 indicates random discrimination, whereas the perfect discriminator has an AUC of 1.

### D. Boosted decision trees

In order to check the discriminative capacity of a combination of HRV features the boosted decision tree classifier was used. Boosting is a method for the creation of an accurate and strong classifier from a set of weak classifiers [15] in a stage-wise fashion, by optimizing an arbitrary differentiable loss function. After a weak learner is trained, the data are then reweighted so that examples that were misclassified at the previous stage now gain weight and examples that are classified correctly lose weight.

Given a set of $N$ training examples of the form $\{(x_1, y_1), \ldots, (x_N, y_N)\}$ such that $x_i$ is the feature vector of the $i$-th example and $y_i$ is its label, the goal is to teach a model, $f$, to predict values in the form, $\hat{y} = \sum_k f_k(x)$. At each stage of gradient boosting, $1 \leq k \leq K$, a model, $f_k$, is constructed. A new improved model is then constructed, $f_{k+1}(x)$, that adds an estimator, $h$ to the previous model: $f_{k+1}(x) = f_k(x) + h(x)$, where $h$ is fitted to the residuals, $y - f_k(x)$.

More formally, the following objective function is minimized:

$$L(\varphi) = \sum_i l(\hat{y}_i, y_i) + \sum_k \Omega(f_k), \quad (1)$$

Here $l$ is a differentiable convex loss function that measures the difference between the prediction and the target. The second term $\Omega$ penalizes the complexity of the regression tree functions. The regularization term helps to smooth the learnt weights to avoid over-fitting.

The extreme gradient boosting classifier, XGBoost [16] is a well-known implementation of boosted decision trees and is widely used in machine learning community, e.g. in Kaggle competitions. The XGB classifier also allows for the retrieval of the importance of each feature as the accumulated gain which is achieved in construction of the ensemble of decision trees.

The epochs of the physiological data are fed into the designed XGB classifier and the probabilities of abnormal outcome are returned for each epoch. The mean probability across the whole recording is then computed to represent the probability of the abnormal outcome – one value per baby.

TABLE I. PREDICTIVE POWER OF HRV FEATURES MEASURED USING AUC. "ORIGINAL" REPRESENTS COMPLETE RECORDINGS FROM 23 SUBJECTS. "HYPOTENSIVE EVENTS" REPRESENT ONLY EPOCHS WITH MAP<GA+4 FROM THE SAME 23 SUBJECTS.

| HRV features | Original (6217 epochs) | Hypotensive events (1488 epochs) | Δ |
|---|---|---|---|
| VLF | 0.67 | 0.7 | ↑ |
| LF | 0.76 | 0.78 | ↑ |
| HF | 0.64 | 0.76 | ↑ |
| LF/HF | 0.73 | 0.69 | ↓ |
| MeanRR | **0.82** | **0.85** | ↑ |
| SDNN | 0.65 | 0.73 | ↑ |
| TINN | 0.7 | 0.77 | ↑ |
| Skewness | 0.58 | 0.52 | ↓ |
| Kurtosis | 0.56 | 0.61 | ↑ |
| ApEn | 0.67 | 0.78 | ↑ |
| SD1 | 0.52 | 0.54 | ↑ |
| SD2 | 0.65 | 0.72 | ↑ |
| RMSSD | **0.76** | **0.87** | ↑ |
| SDNN/RMSSD | 0.64 | 0.65 | ↑ |
| AllanFactor | 0.51 | 0.55 | ↑ |

Δ represents an increase or decrease of the performance.

### E. Model selection and performance assessment

In this work the performance of the classifier was similarly measured by AUC. The leave-one-out (LOO) subject independent performance assessment is used in this work. All but one subject's data were used for training and the remaining subject's data were used for testing. The procedure was repeated until each patient had been a test subject.

The user-tuneable parameters of the XGB classifier were selected using the 5 times 2-fold cross validation (CV) which was performed on the training data. CV folds were designed in a way that preserved the subject integrity. This allows for a model selection routine that has an optimization criteria that matches the one of the LOO performance assessment routine – subject-independent evaluation.

### III. RESULTS AND DISCUSSION

The results of the predictive power of HRV are presented in Table I. The AUC was evaluated for different subsets of HRV features – using the whole recording or using only those segments where MAP < GA+4.

It can be seen from Table I that epoch-based features bear moderate predictive power with respect to the outcome, with mean RR reaching an AUC of 0.82. Importantly, examining HRV during episodes of low blood pressure improves the predictive power of almost every HRV feature, with RMSSD achieving the highest discriminating power with an AUC of 0.8 7, a substantial improvement from 0.76.

It is worth noting that in both cases the same 23 preterms are used. Newborns without hypotensive episodes were excluded from the experiment as explained in Section II.A. This ensures that the two corpuses – original dataset and hypotensive events – are comparable and the improvement is purely attributable to the effect that BP has on HRV rather than an effect which could be caused by including extra subjects in the original dataset category (those newborns whose MAP ≥GA+4).

Fig. 1 shows the probability density function of the RMSSD feature for the original dataset, the subset with normal BP and the subset of hypotensive events, for the cohort of preterms with both healthy and unhealthy outcomes. From Fig. 1 it can be seen that the HRV, in terms of the energy of the successive differences between neighbouring RR intervals, is generally higher for the healthy cohort than unhealthy for both the whole recording and for the episodes with normal BP, with a substantial overlap between the two distributions. From Fig. 1 (c) it can be seen that the separation increases when considering only hypotensive episodes. If we compare distributions of RMSSD for the cohort of preterms with the healthy outcome, for episodes with normal BP (Fig.1 (b)) and low BP (Fig. 1 (c)), it can be seen that HRV increases with the decreased BP. At the same time, for the cohort with unhealthy outcome, the HRV stays the same regardless of the BP values. The healthier preterms react to the change in BP.

In [17] the increased level of interaction between EEG and BP in preterms was shown to be associated with lower risks of illness severity. Similarly here as shown in Fig. 1, the increased "interaction" between HRV and BP is associated with good outcome, whereas the lack of response is associated with poor outcome.

Fig. 2 shows the 3D projection of the principal component analysis applied to the set of all features on the original dataset and hypotensive subset. The PCA projection of the combination of the MeanRR and fourteen HRV features on the original dataset results in no clear discrimination between healthy and unhealthy neonates whereas a clear separation is observable for the hypotensive subset when the features are combined.

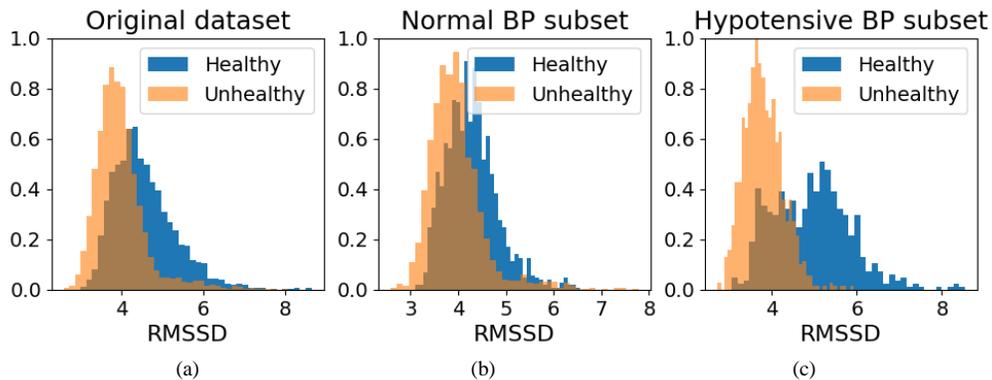

(a) (b) (c)

Figure 1. PDF for the RMSSD feature. Original subset (a) contains RMSSD feature values from the complete recordings. Normal (b) and hypotensive (c) subsets represent RMSSD feature extracted during episodes of normal BP ($MAP \geq GA + 4$) and during hypotensive events ($MAP < GA + 4$).

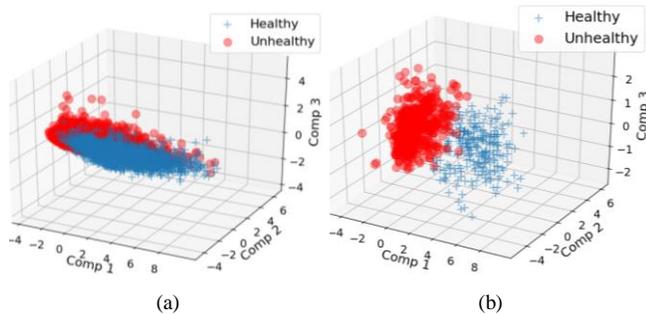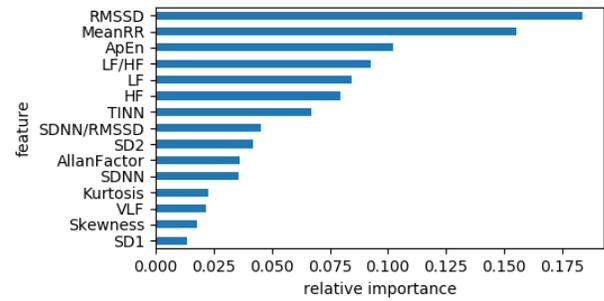

Figure 2. Principal component analysis (PCA) of the original dataset (a) and its hypotensive subset (MAP<GA+4) (b). In this study PCA is used as a tool for exploratory feature analysis aimed at checking the discriminative power of HRV features with respect to the short-term outcome of the preterm neonate.

Figure 3. Mean of feature importance reported by the boosted decision trees classifier.

Feeding all fifteen features extracted for hypotensive subset into a boosted decision tree classifier further improved the performance, reaching an AUC of 0.97, in comparison to an AUC of 87% from the single best feature, RMSSD. This indicates that by using multimodal data, HRV and BP, an objective decision support tool for clinical prediction of short-term outcome in preterms who suffer episodes of low BP can be built. The tool outputs a probability of good outcome for each 5 minute window and can guide and assist the healthcare professional in critical decision making.

Fig. 3 plots the importance of the various features which is computed as a by-product of training the XGB classifier. The importance is computed as an average across 23 iterations in the LOO procedure. It can be seen that RMSSD and MeanRR are the two most important features for the short-term outcome prediction. RMSSD was previously shown to improve 2 year outcome prediction for preterm neonates [7]. MeanRR, a key time domain feature that is the inverse of the heart rate, is used for derivation of a number of HRV metrics. The boosted decision tree classifier has discovered that the combination of the heart rate and HRV is important for accurate prediction of the outcome.

This is the first study which investigates an association between HRV and neonatal health outcome in the context of episodes of low BP. Obtained results indicate that HRV is sensitive to BP and features extracted during the episodes of low MAP improve the prediction of short-term outcome for the preterm neonates. ECG and BP records are usually available soon after birth and this work presents a promising step towards the use of multimodal data in building an objective decision support tool for the clinical prediction of short-term outcome in preterms who suffer from hypotension. Future research will concentrate on the incorporation of other modalities, such as EEG, for both short-term and long-term outcome prediction.